\documentclass[referee]{aa} % for a referee version

\newcommand{\ltsim}{\protect\raisebox{-0.5ex}{$\:\stackrel{\textstyle <}{\sim}\:$}}

\usepackage{graphicx}
\usepackage[round]{natbib}

\begin{document}

\title{{Deriving temperature, mass and age of evolved stars from high-resolution spectra}
\subtitle{Application to field stars and the open cluster IC 4651 }
\thanks{Based on observations collected at the ESO telescopes at the Paranal and La Silla Observatories, Chile.}}
   
\author{ K. Biazzo\inst{1,2} \and L. Pasquini\inst{2} \and  L. Girardi\inst{3} 
\and  A. Frasca\inst{1} \and L. da Silva\inst{4} \and J. Setiawan\inst{5} \and E. Marilli\inst{1} 
\and A. P. Hatzes\inst{6} \and S. Catalano\inst{1}} 

\offprints{K. Biazzo, \email{katia.biazzo@oact.inaf.it}}

\institute{INAF - Osservatorio Astrofisico di Catania, via S. Sofia 78, 95123 Catania, Italy
\and ESO - European Southern Observatory, Karl-Schwarzschild-Str. 3, 85748 Garching bei M\"unchen, Germany
\and INAF - Osservatorio Astronomico di Padova, vicolo dell'Osservatorio 5, 35122 Padova, Italy
\and Observat\'orio Nacional, R. Gal. Jos\'e Cristino 77, 20921-400 S\~ao Crist\'ov\~ao, Rio de Janeiro, Brazil
\and Max Planck-Institute f\"ur Astronomie, Heidelberg,  K\"onigstuhl 17, 69117 Heidelberg, Germany
\and Th\"uringer Landessternwarte, Tauterburg, Sternwarte 5, 07778 Tauterburg, Germany}

\date{Received / accepted}

\abstract
{}
{We test our capability of deriving stellar physical parameters of giant stars by analysing a sample of 
field stars and the well studied open cluster IC 4651 with different spectroscopic methods.}
{The use of a technique based on line-depth ratios (LDRs) allows us to determine with high precision the 
effective temperature of the stars and to compare the results with those obtained with a classical LTE 
abundance analysis.}
{($i$) For the field stars we find that the temperatures derived by means of the LDR method are in excellent 
agreement with those found by the spectral synthesis. This result is extremely encouraging because it shows that 
spectra can be used to firmly derive population characteristics (e.g., mass and age) of the 
observed stars. ($ii$) For the IC 4651 stars we use the determined effective temperature to derive the following 
results. a) The reddening $E(B-V)$ of the cluster is $0.12\pm0.02$, largely
independent of the color-temperature calibration used. b) The age of the cluster is $1.2\pm0.2$ Gyr. 
c) The typical mass of the analysed giant stars is $2.0\pm0.2M_{\sun}$. 
Moreover, we find a systematic difference of about 0.2 dex in $\log g$ between spectroscopic and evolutionary values.}
{We conclude that, in spite of known limitations, a classical spectroscopic analysis of giant stars may 
indeed result in very reliable stellar parameters. We caution that the quality of the agreement, on the 
other hand, depends on the details of the adopted spectroscopic analysis.}
\keywords{Stars: late-type  --  Galaxy: open cluster and associations: individual: IC 4651  --  Techniques: 
spectroscopic}
\titlerunning{Temperature, mass and age of evolved stars}
\authorrunning{K. Biazzo et al.}
\maketitle

\section{Introduction}

The detailed study of stellar populations in our own Galaxy and in its neighborhoods has received a major impulse 
in the last years, thanks to the use of large telescopes coupled to multi-object high-resolution spectrographs. 

The spectroscopic determination of chemical abundances in the atmospheres of stars may greatly contribute to our 
knowledge of galaxies. In fact, once a set of chemical abundances is complemented with a stellar age, it is possible 
to assess the age-metallicity relation, and to invert the observed color-magnitude diagram (CMD) thus obtaining the Star 
Formation History (SFH) of the Galaxy (see, e.g., \citealt{Tolstoy05}, and the Large Magellanic Cloud case illustrated 
by \citealt{Cole05}). When individual stellar ages are unknown, this analysis becomes more uncertain because 
of the so-called age-metallicity degeneracy, i.e. the fact that old metal-poor stars can occupy the same region of 
the CMD as young metal-rich objects.

In this context, open clusters provide fundamental tools because each cluster represents an homogeneous sample of stars 
having the same age and chemical composition. Moreover, they are very suitable for the investigation of several issues 
related to  stellar and Galactic formation and evolution. In particular, young open clusters provide information about 
present-day star formation processes and are key objects for clarifying questions on galactic structure, while old and 
intermediate-age open clusters play an important role in linking the theories of stellar and galactic evolution. 

The main classical tool to study cluster properties is the color-magnitude diagram, which suffers from
several uncertainties and intrinsic biases, such as the limited knowledge of the chemical composition 
of the stars and the degeneracy in deriving the distance (and therefore age) and reddening. As a consequence, 
the photometric analysis of open clusters alone might not be so conclusive for an accurate determination  of
ages, distances, metallicities, masses, color excesses, and temperatures, as shown by 
\cite{Randich05}. Thus, spectroscopic methods to determine the effective temperature of cluster members 
are efficient techniques that are independent of the cluster reddening. This reddening can then be
obtained by comparing the spectroscopic results to the photometric ones.

Effective temperatures can be determined by imposing the condition that the derived abundance 
for one chemical element with many lines in the spectrum (typically \ion{Fe}{i}) does not depend on 
the excitation potentials of the lines (hereafter this technique will be called as ``spectral synthesis'' and 
the ``spectroscopic temperature'' derived in this way will be indicated $T_\mathrm{eff}^{\rm SPEC}$). 
Another spectroscopic method is based on the ratio of the depths of two lines having different sensitivity to 
effective temperature. This line-depth ratios (LDRs) technique provides an excellent measure of stellar temperature 
(hereafter $T_\mathrm{eff}^{\rm LDR}$) with a sensitivity as small as a few Kelvin degrees in the most favorable 
cases (\citealt{Gray91,Stras00,Gray01}).

In this paper we apply the LDR method to derive effective temperature of nearby evolved field stars 
with good Hipparcos parallaxes (accuracy better than 10\%, \citealt{daSil06}) 
and of giant stars of the intermediate-age open cluster IC 4651. For both 
groups, the temperature was previously derived photometrically by means of color indices and 
spectroscopically by abundance studies (\citealt{Pasqu04,daSil06}). In this work, we compare $T_\mathrm{eff}^{\rm LDR}$ 
with the previous photometric and spectroscopic determinations in order to check temperatures obtained with different 
methods, and to understand the temperature range in which each method can be used. This will verify how accurately 
physical parameters can be obtained by inverting the spectroscopic results. Moreover, for the stars belonging to the 
open cluster IC 4651, we are able to derive in a ``spectroscopic way'' color excess, mass and age. This is an important 
test because, to our knowledge, this is the first time that these parameters are determined in a cluster by 
means of the line-depth ratio method. Moreover, the possibility of comparing our spectroscopic results 
on IC 4651 with those obtained by the classical fitting of the main sequence provides a unique opportunity to 
cross-check our inversion method.

\section{Sample selections and observations}

We have considered seventy-one evolved field stars and six giant stars belonging to the intermediate-age open cluster 
IC 4651. 

The sample of field stars has been selected and analysed by \cite{Setia04} to derive the radial velocity (RV) 
variations along the Red Giant Branch and to investigate the nature of the radial velocity long-term variations. 
Subsequently, the same sample was studied by \cite{daSil06} for the determination of radii, temperatures, masses, 
and chemical composition with the aim to understand how the RV variability is related to the stellar 
physical parameters. 

\begin{figure}
\centering
\includegraphics[width=9cm]{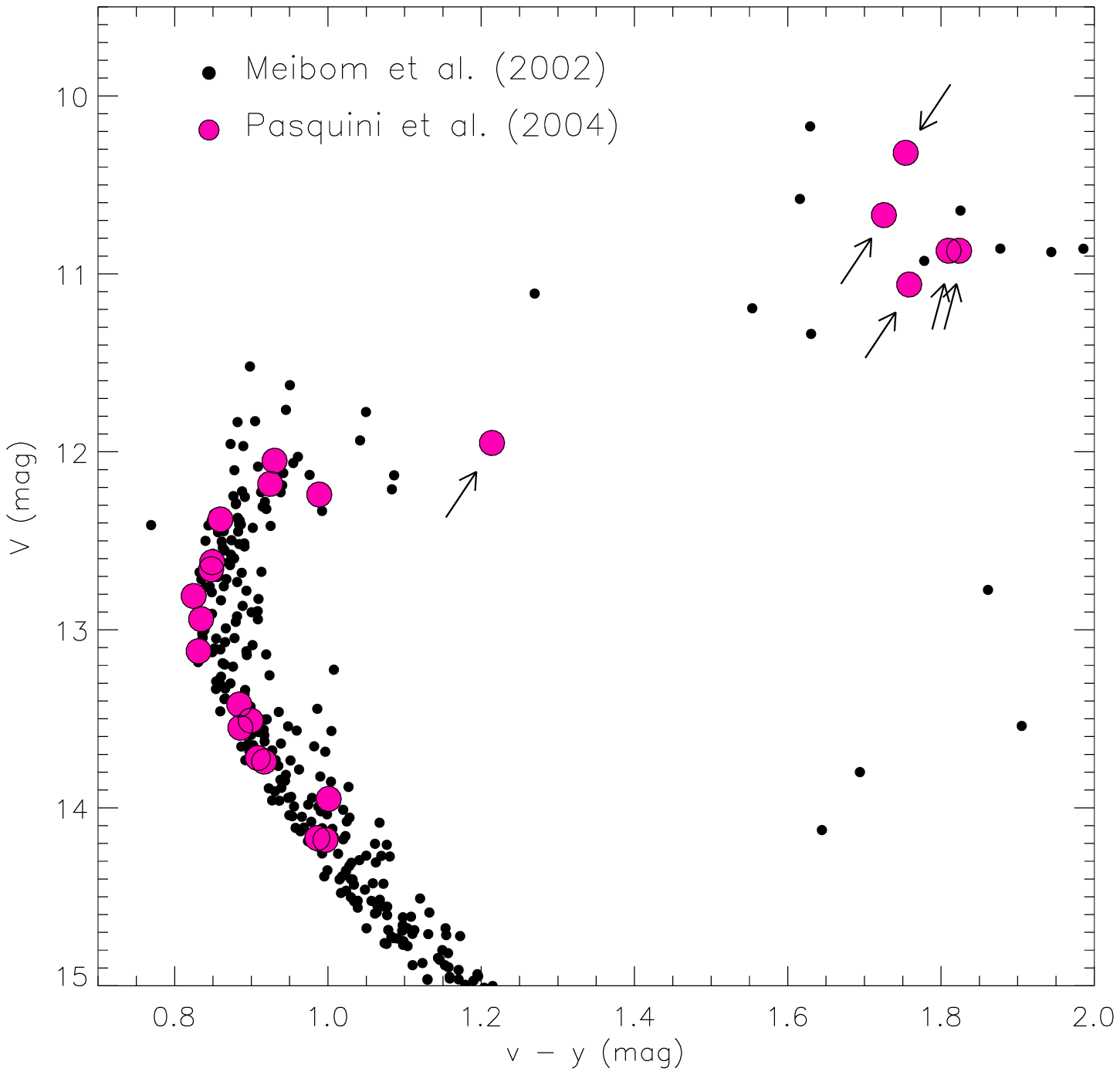} 
%\vspace{-.6cm}
\caption{CMD of IC~4651. $V$ and $v-y$ are taken from \cite{Anth88} and \cite{Meibom02}, respectively. The stars 
analysed by \cite{Meibom02} and \cite{Pasqu04} are indicated by different symbols, while our ones are marked 
with arrows.}
\label{fig:cmd_ic4651}
\end{figure}

The stars in the cluster IC 4651 have been taken from the sample of \cite{Pasqu04}. The evolutionary status 
of the analysed stars is indicated by their position in the CMD shown in Fig.~\ref{fig:cmd_ic4651} in which our 
stellar sample is marked with arrows. 

The details of the observations, data reduction, and instrumentation are described 
in \cite{Setia04} and \cite{Pasqu04}. Their main characteristics are briefly summarized here. 
The spectra of the field stars were acquired with the FEROS 
spectrograph ($R=48\,000$, \citealt{Kaufer1999}) at the ESO 1.5m-telescope in La Silla (Chile), while the IC 4651 
spectra were acquired with the UVES spectrograph ($R=100\,000$, \citealt{Dekker2000}) at the ESO VLT Kueyen 8.2m-telescope 
in Cerro Paranal (Chile). The FEROS spectrograph has a wavelength coverage between 3700 and 9200 \AA, while the spectral 
region of the UVES observations covers the 5800$-$6800 \AA~range (we have selected the data acquired by 
\citealt{Pasqu04} with the CD\#3 since they covered the spectral region of our interest). The signal-to-noise ($S/N$) 
ratio in all cases is higher than 150/pixel and 160/pixel for the FEROS and UVES spectra, respectively.

%--------------------------------------------------------------------------------------------------------------------
%--------------------------------------------------------------------------------------------------------------------

\section{Data analysis}

\subsection{Effective temperature}

The spectral regions covered by FEROS and UVES contain a series of weak metal lines which can be used 
for temperature determination with the LDR method. Lines from similar elements such as iron, vanadium, titanium, but 
with different excitation potentials ($\chi$) have different sensitivity to temperature. This is because 
the line strength is a function of temperature and, to a lesser extent, of the electron pressure (and 
consequently, the surface gravity). This 
sensitivity arises from the exponential and power dependence on temperature in the excitation and ionization processes. 
Different lines are formed at different depths of the stellar atmosphere and therefore one would expect that lines 
from levels with high excitation potentials are formed deeper in the atmosphere and hence at higher temperatures. 
For this reason, it is better to choose pairs of lines with the largest $\chi$-difference. Between 6180 
\AA~and 6280 \AA~there are several lines of this type whose depth ratios have been exploited for temperature 
calibrations (\citealt{Gray91,Gray01,Cata02,Biazzo07a}), to study the rotational modulation of the average effective 
temperature of magnetically active stars (\citealt{Frasca05, Biazzo07b}) or to investigate in Cepheid stars the 
pulsational variations with phases (\citealt{Kovty00, Biazzo04}). In particular, we chose 15 weak line pairs 
for which \cite{Biazzo07a} made suitable calibrations. These lines are of elements belonging to the iron group 
that have different temperature sensitivity. Furthermore, the $S/N$ ratio around 150 for all the spectra makes them 
very suitable for the LDR method, because with such a high $S/N$ ratio a precision of a few Kelvin degree is expected 
(\citealt{Gray91,Gray01,Kovty06,Biazzo07a}).

\subsubsection{Sample of field stars}

All the field stars analysed in this paper have low rotational velocity ($v\sin i < 7$ km s$^{-1}$). 
The spectral resolution of FEROS is quite similar to that of the spectrograph ELODIE ({\it Observatoire de Haute-Provence}, 
France), used by \cite{Biazzo07a}. As a consequence, the LDR$-T_\mathrm{eff}$ calibrations developed by \cite{Biazzo07a} 
at $v\sin i \le 5$ km s$^{-1}$ have 
been applied to derive the effective temperature ($T_\mathrm{eff}^{\rm LDR}$). Our star sample is comprised of 
sixty-seven giants and four stars with log $g>$ 4.0 (namely HD2151, HD16417, HD26923 and HD62644), as derived by 
\cite{daSil06}. For the latter the LDR$-T_\mathrm{eff}$ calibration for main sequence stars obtained 
by \cite{Biazzo07a} has been used.

We would like to note that \cite{Biazzo07a} adopted the color-temperature relationship obtained by \cite{Gray05}, 
which does not include corrections due to metallicity. We shall come back to this point in the discussion of the Sections 
\ref{sec:evol_parameters_teff} and \ref{sec:IC4651_teff}.

\subsubsection{IC 4651}

For the giant stars of IC 4651, the UVES spectra have $R=100\,000$, significantly higher than the ELODIE ($R=42\,000$), so, 
in order to apply the calibrations, it was necessary to degrade the UVES spectra to the resolution of ELODIE. 
The analysed six giant stars have rotational velocities $v\sin i$ in the range 
0--15 km s$^{-1}$, thus the LDR$-T_\mathrm{eff}$ calibrations developed by \cite{Biazzo07a} for the appropriate $v\sin i$ 
have been applied to derive the effective temperatures ($T_\mathrm{eff}^{\rm LDR}$). For the subgiant E95, an interpolation 
between the main sequence and the giant calibrations was applied. Consequently, the effective temperature uncertainty for 
this star should be higher than for the giants.

\subsection{Surface gravity}
\label{sec:surf_grav}
We can compute the ``photometric'' surface gravity using the relation 
$\log g= \log g_{\sun} + \log (M/M_{\sun}) + 4 \log (T_\mathrm{eff}/T_{\rm eff, \sun}) - \log (L/L_{\sun})$, where 
$g_{\sun}$ is the solar surface gravity, while $M$, $T_\mathrm{eff}$ and $L$ are the 
stellar mass, photometric effective temperature, and luminosity in the respective solar units.

\subsection{Reddening}

Several determinations of the photometric reddening of IC 4651 exist in literature (see, e.g., \citealt{Meibom02}, and 
reference therein). Once the spectroscopic temperatures have been derived, we can estimate for each star the 
intrinsic color index ($B-V$)$_0$ by inverting the $(B-V)_0-T_\mathrm{eff}$ relation and to compute the 
color excess $E(B-V)$ from the observed and reddened $(B-V)$ color. This can be obtained for each star separately, 
therefore the statistics on all stars can tell us not only the reddening of the cluster, but also provide a 
check of the goodness of the method applied.

\subsection{Mass and Age}

Mass and age are estimated using a slightly modified version of the Bayesian estimation method 
conceived by \cite{Jorg05}, as \cite{daSil06} proposed (see references therein 
for more information). In short, given the stellar absolute magnitude, effective temperature,
metallicity, and the associate errors, we can estimate the probability that such a star belongs to each 
small section of a theoretical stellar isochrone 
of a given age and metallicity. In particular, we have considered the isochrones developed by \cite{Gira00}. Then, 
the probabilities are summed over the complete isochrone, and 
hence over all possible isochrones, by assuming a Gaussian probability of having the observed metallicity and its error, 
and a constant probability of having stars of all ages. The latter assumption is equivalent to assuming a constant star 
formation rate in the solar neighbourhood. In this way, at the end, we have the age probability distribution function (PDF) 
of each observed star. PDFs can also be obtained for any stellar property, such as initial mass, surface gravity, intrinsic 
colour, etc\footnote{A Web version of this method is available at the URL 
http://stev.oapd.inaf.it/$^\sim$lgirardi/cgi-bin/param.}.

Although a full discussion of the PDF method is beyond the scope of this paper, we note the following. 
The method, with just some small differences, has already been tested on both main sequence stars (\citealt{Nordstrom04}) 
and on giants and subgiants (\citealt{daSil06}). Ages of dwarfs turn out to be largely undetermined by this method, 
due to their very slow evolution while on the main sequence. Ages of giants turn out to be well determined provided 
that the effective temperature and the parallax (absolute magnitude) are measured with enough accuracy. In fact, 
\cite{daSil06} find that stars with errors of 70~K in $T_{\rm eff}$, and less than 10\% errors in parallaxes, have ages 
determined with an accuracy of about 20\%. These errors become larger for particular regions of the CMD, for instance on 
the red clump region, where stars of very different age and metallicity become tightly clumped together, and where in 
addition there is a superposition of red clump stars and first-ascent RGB ones. On the other hand, the best age 
determinations are expected on the subgiant branch region of the CMD, where evolutionary tracks of different masses 
separate very well from each other. To some extent, also the lower part of the red giant branch provides good age 
determinations, since in this CMD region there is no superposition with other evolutionary stages and stars above a given 
absolute magnitude can only be fitted by stars below a given mass (and above a given age).

Of course, the above-mentioned errors do not include the systematic errors that may hide in the stellar evolutionary 
tracks and isochrones used. In fact, in any set of evolutionary tracks the position of the giants depends on the choice 
of mixing length parameter (usually calibrated on the Sun), and to a much smaller extent on details of solving the 
atmosphere structure and on the interpolation of opacity tables. The only way to avoid this kind of uncertainty would be 
the building of tracks (and isochrones) in which the theory of energy transport and atmospheric structure are accurately 
calibrated on observations of star clusters and binaries. This approach however is still too far to reach. Using the 
PDF method with different sources of evolutionary tracks could provide hints on the magnitude of systematic errors, but 
would not solve the situation because different sets of tracks in the literature do share the same assumptions and 
input physics. 

Therefore, at present we cannot reliably evaluate the systematic errors present in the PDF method. A simple check with two Hyades 
giants, where the PDF ages can be compared with the turn-off ones, hints to an accuracy of about 20\% for 
the objects studied by \cite{daSil06}. It would be extremely interesting to perform similar tests using giants in clusters of 
a wide range of ages, although we know that their distances (and turn-off ages) would be significantly more uncertain than 
the Hyades ones.

\section{Results}
\subsection{Sample of field stars}

\subsubsection{Effective temperature}
\label{sec:evol_parameters_teff}

In Table \ref{tab:evol_parameters} we list names, $T_\mathrm{eff}^{\rm LDR}$ together with the photometric 
($T_\mathrm{eff}^{\rm PHOT}$) and spectroscopic ($T_\mathrm{eff}^{\rm SPEC}$) temperatures, and as well as 
the stellar abundances retrieved by \cite{daSil06}. The photometric temperatures were determined using the 
$(B-V)-T_\mathrm{eff}$ relationships of \cite{Alonso96, Alonso99} for dwarf and giant stars.

\begin{figure*}
\centering
\includegraphics[width=8.9cm]{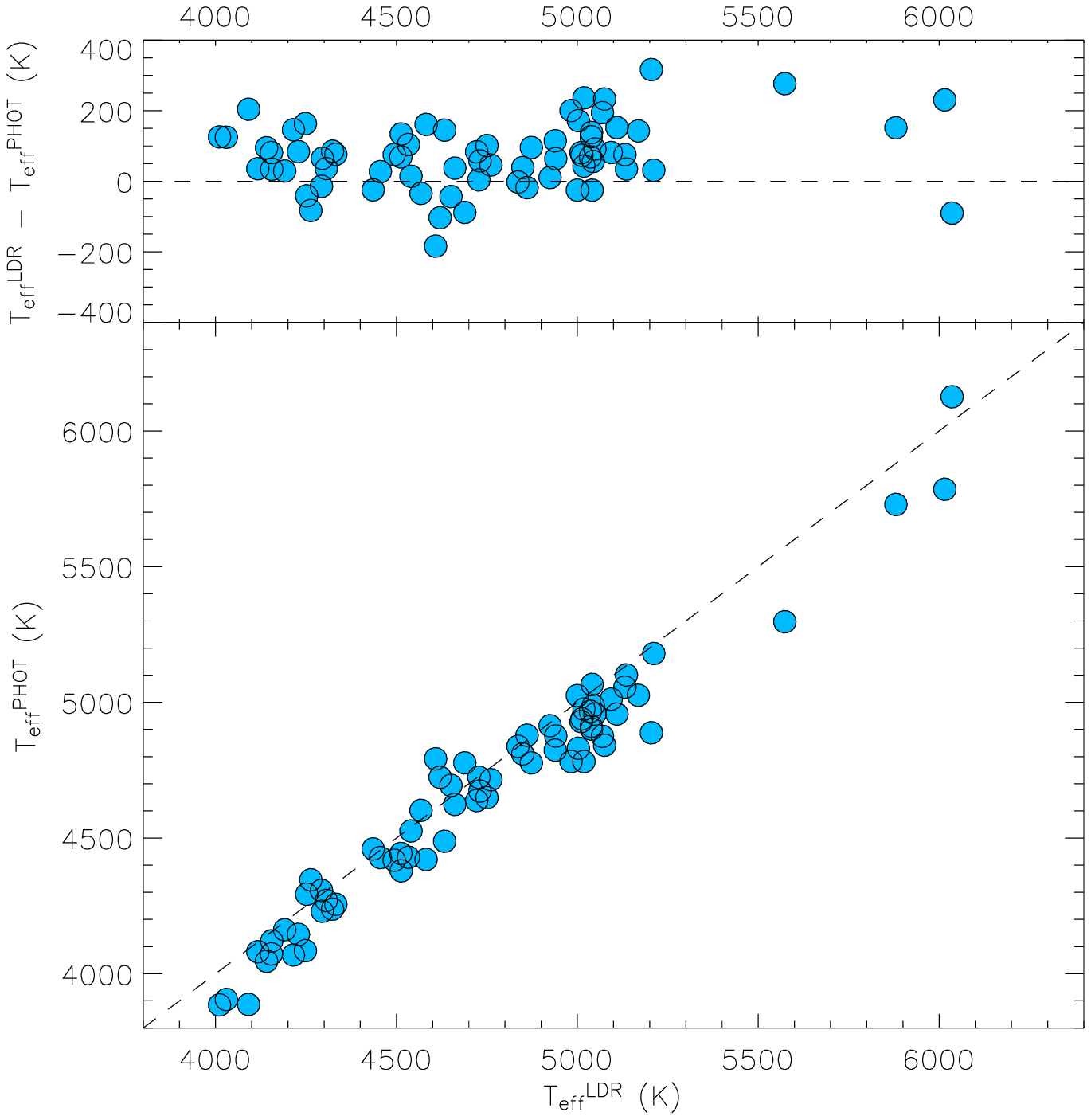}
\includegraphics[width=8.9cm]{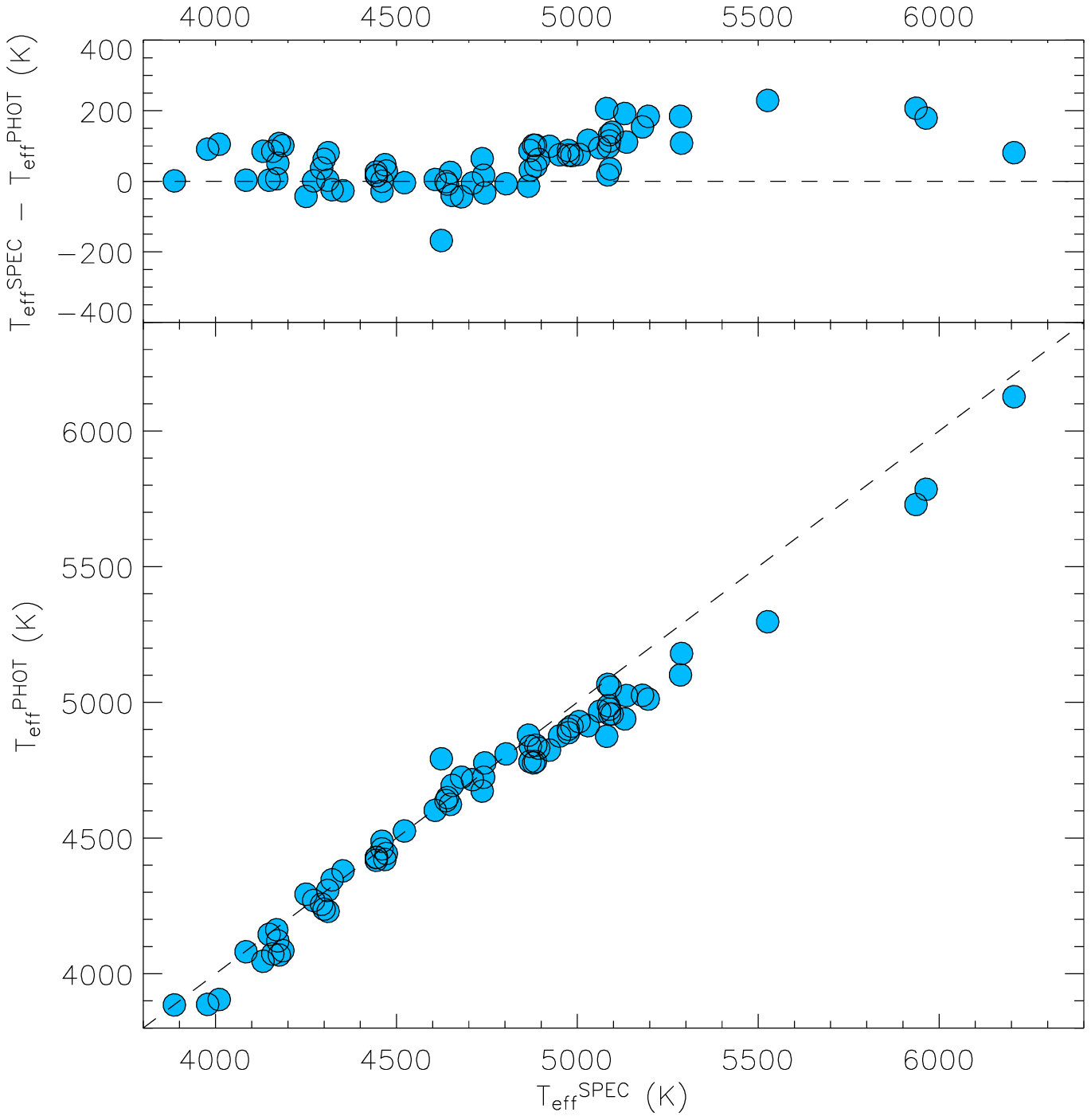}
\includegraphics[width=8.9cm]{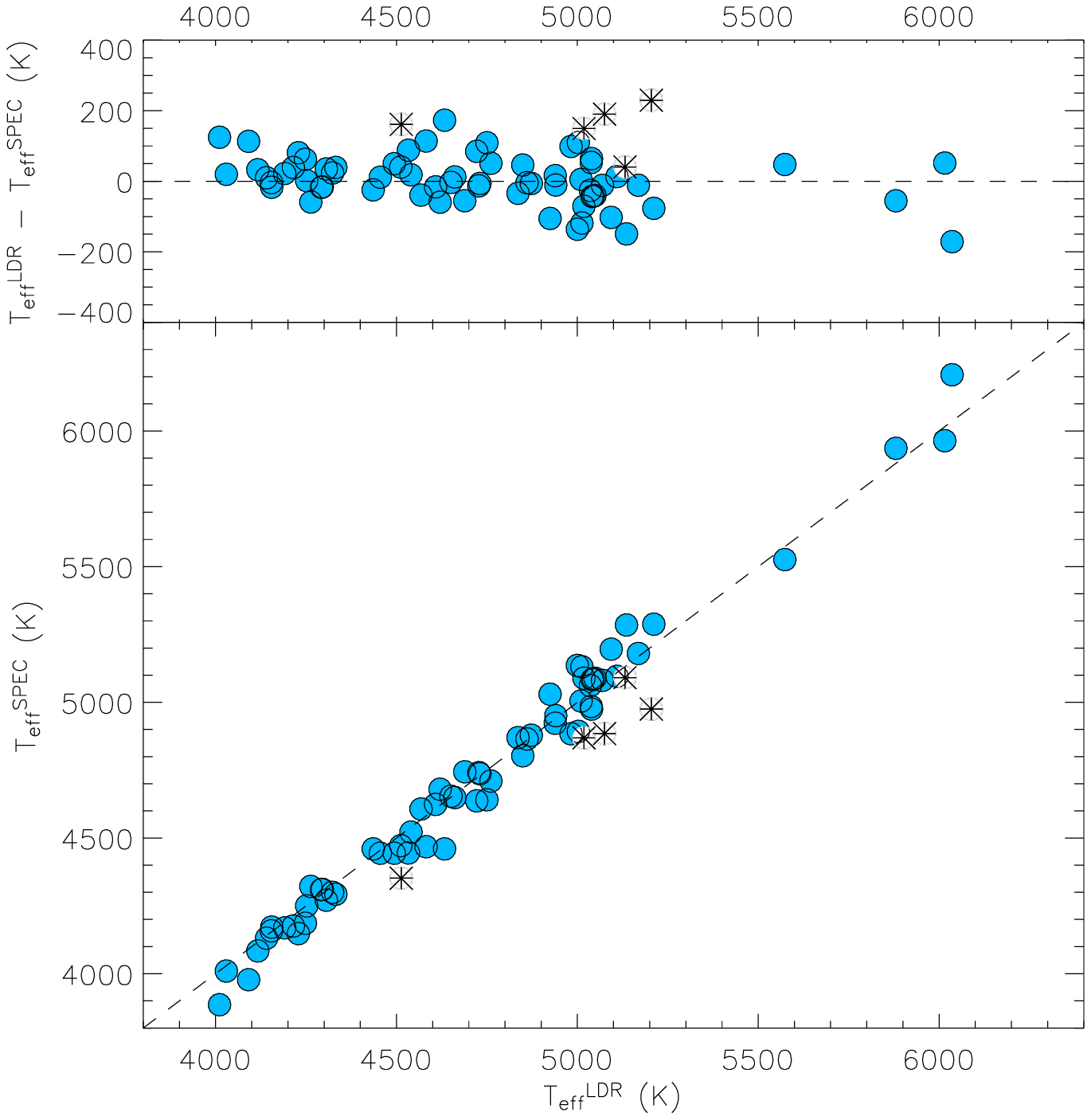} 
\vspace{-.8cm}
\caption{Comparison between effective temperatures obtained by line-depth ratio, spectral synthesis, and 
photometric methods. The second plot ($T_\mathrm{eff}^{\rm PHOT}$ vs. $T_\mathrm{eff}^{\rm SPEC}$) shows the results 
obtained by \cite{daSil06}. The asterisks in the third plot represent the five metal-deficient ([Fe/H]$<-0.4$) 
stars in our sample.}
\label{fig:evol_teff_comparison}
\end{figure*}

\begin{figure}
\centering
\includegraphics[width=9cm]{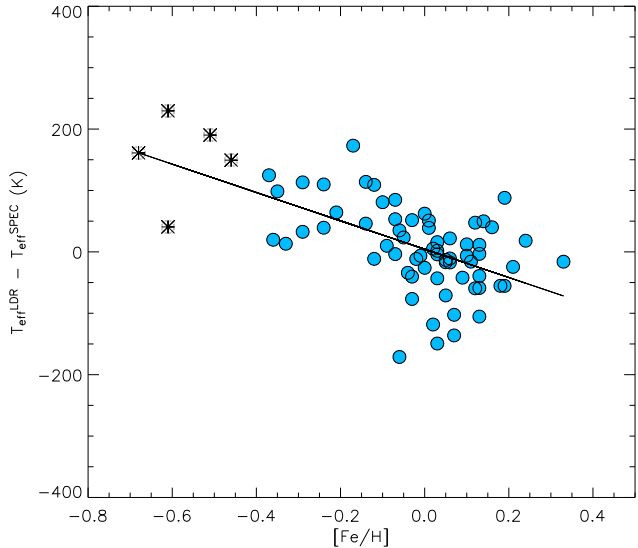}
\caption{Metallicity dependence of $T_\mathrm{eff}^{\rm LDR}-T_\mathrm{eff}^{\rm SPEC}$. The five stars with [Fe/H]$<-$0.4 
are showed with asterisks. Continuous line is the linear fit to the points.}
\label{fig:evol_teff_met_depend}
\end{figure}

The comparison between these three temperatures is emphasized in Fig.~\ref{fig:evol_teff_comparison} and discussed 
here in the following. 
\begin{itemize}
\item[i.] The temperatures derived by spectral synthesis ($T_\mathrm{eff}^{\rm SPEC}$) and by 
LDR method ($T_\mathrm{eff}^{\rm LDR}$) are higher than the photometric ones by about 60--70~K, on average. 
This indicates that the photometric techniques tend to underestimate the temperature compared to the 
spectroscopic ones. This result seems to be more evident for the hottest stars in our sample. 
\item[ii.] The agreement between $T_\mathrm{eff}^{\rm PHOT}$ and the $T_\mathrm{eff}^{\rm SPEC}$ is better in 
the temperature range 4000--5300~K, as already found by \cite{daSil06} and displayed in the top-right panel of 
Fig.~\ref{fig:evol_teff_comparison}. The good agreement between $T_\mathrm{eff}^{\rm PHOT}$ and $T_\mathrm{eff}^{\rm SPEC}$ 
or $T_\mathrm{eff}^{\rm LDR}$ in the 4000--5300~K temperature range is likely due to the fact that all these field stars 
are nearby and, consequently, the reddening, which is the parameter that mainly influences the photometric effective 
temperature determination, is negligible. The temperatures derived by spectroscopic methods are instead 
reddening-free. At higher temperatures, the behaviour of $T_\mathrm{eff}^{\rm PHOT}$ versus $T_\mathrm{eff}^{\rm SPEC}$ or 
$T_\mathrm{eff}^{\rm LDR}$ reflects the shape of the photometric calibration used for the temperature 
determination.
\item[iii.] The temperatures obtained by means of the LDR method are in very good agreement in all 
the range with the spectroscopic values computed by \cite{daSil06}, the average difference 
$<T_\mathrm{eff}^{\rm LDR}-T_\mathrm{eff}^{\rm SPEC}>$ being about 15~K with rms = 78~K. 
This rms can be considered as an estimate of the absolute error on the star temperature. 
%, but the temperature difference among stars is much better defined.
\item[iv.] The $T_\mathrm{eff}^{\rm PHOT}$ versus $T_\mathrm{eff}^{\rm LDR}$ plot shows larger scatter compared to 
$T_\mathrm{eff}^{\rm PHOT}$ versus $T_\mathrm{eff}^{\rm SPEC}$. If we consider the whole temperature range, we obtain 
$<T_\mathrm{eff}^{\rm LDR}-T_\mathrm{eff}^{\rm PHOT}>=73~\mathrm{K}$ with the rms of 92~$\mathrm{K}$. The mean difference 
$<T_\mathrm{eff}^{\rm SPEC}-T_\mathrm{eff}^{\rm PHOT}>$ is $58~\mathrm{K}$ and the rms is 72~K. The fact that 
$T_\mathrm{eff}^{\rm PHOT}$ vs. $T_\mathrm{eff}^{\rm LDR}$ shows higher scatter compared to $T_\mathrm{eff}^{\rm PHOT}$ 
vs. $T_\mathrm{eff}^{\rm SPEC}$ can be due to the slight dependence of $T_\mathrm{eff}^{\rm LDR}$ on metallicity, 
especially for stars with [Fe/H] significantly different than zero, as pointed out by \cite{Biazzo07a}. In fact, 
if we plot the difference $T_\mathrm{eff}^{\rm LDR}-T_\mathrm{eff}^{\rm SPEC}$ as a function of the metallicity 
(Fig.~\ref{fig:evol_teff_met_depend}), a residual dependence of the effective temperature on the metallicity emerges, 
mainly due to the metal-poor stars. If we disregard the five metal-deficient stars with [Fe/H]$<-$0.4 (asterisks in 
Fig.~\ref{fig:evol_teff_comparison}), the average difference $<T_\mathrm{eff}^{\rm LDR}-T_\mathrm{eff}^{\rm SPEC}>$ 
becomes very low (5~K) with a value of 68~K as rms. If we instead take the linear relationship (full line in 
Fig.~\ref{fig:evol_teff_met_depend}) and correct $T_\mathrm{eff}^{\rm LDR}$ for the residual dependence on the 
metallicity, the final residuals between the two temperatures $T_\mathrm{eff}^{\rm LDR}$ and $T_\mathrm{eff}^{\rm SPEC}$ 
is only 48~K. The agreement is remarkably good considering that in this small scatter are included all the possible 
differences, such as errors in the equivalent width and in the line-depth ratio measurements, errors in the 
LDR--$T_\mathrm{eff}$ calibrations and errors in the $T_\mathrm{eff}$ scale of the adopted standard stars. We think that 
the residual dependence of $<T_\mathrm{eff}^{\rm LDR}-T_\mathrm{eff}^{\rm SPEC}>$ on the metallicity is due to the 
effect of [Fe/H] on LDRs that is a ``second order'' one compared to temperature and gravity, but can be neglected 
only for stars with a near solar metallicity. In order to adequately take into account this effect, a proper 
LDR$-T_\mathrm{eff}$ calibration should be done for a large sample containing many stars with well determined 
metallicity, spanning a wide range. 
\end{itemize}

One could object that the fact the two spectroscopic temperatures ($T_\mathrm{eff}^{\rm LDR}$ 
and $T_\mathrm{eff}^{\rm SPEC}$) are in very good agreement should be expected, 
since the methods are very similar. However, there are a number of important differences that one should not,
{\it a priori}, expect such a good agreement.
\begin{itemize}
\item The two techniques are independent. One method simply measures several LDRs and takes advantage of the 
LDR--temperature calibrations of standard stars. These calibrations use a $(B-V)-T_\mathrm{eff}$ polynomial relationship 
to determine the temperature of the standard stars. The other method computes the temperature from the equivalent 
widths eliminating any dependence of the \ion{Fe}{i} abundance on the line excitation potential and using appropriate 
atmospheric models. In order to compute the \ion{Fe}{i} abundance, the process involves the use of stellar atmospheric 
models, the use of line strength ($\log g f$), the determination of microturbulence ($\xi$) and gravity, a process which 
is not present in the LDR method. Only when all these steps are properly made, the results are similar.
\item We know that the lines are formed at different stellar atmospheric levels, but if one chooses pairs
of weak lines of similar elements (such as elements of the iron group), the abundance dependence is practically 
eliminated when the ratio between the two lines is considered. In the measure of the equivalent width, the abundance 
dependence is instead always present.  
\item The effects of macroturbulence cancel out, to the first order, when the LDR is computed because they affect all 
lines in the same way. These effects are not present in the equivalent width measurements.
\item The continuum choice influences in a strong way the equivalent width determination. As a first approximation, 
this does not happen for the LDRs, because the continuum is practically the same for each line pair which are 
very close in wavelength. 
\item The rotational velocity shows its effect in the line-depth ratio computation, but can be taken into account using 
appropriate calibrations at proper rotational velocity (\citealt{Biazzo07a}). The equivalent width 
measurement is not instead affected by rotational broadening (\citealt{Gray05}).
\item Microturbulence is negligible when the depth ratio between a line pair is 
considered. The effective temperatures derived by spectral synthesis are instead obtained computing the 
contribution of microturbulence. As a consequence, wrong microturbulence values (also of 0.1-0.2 km s$^{-1}$) leads to 
wrong temperature determinations.
\end{itemize}
Given all these differences, there are no {\it a priori} reasons why the effective temperatures resulting from the 
spectral synthesis and LDR methods had to be similar.

\begin{table}
\caption{Parameters of the field stars.\label{tab:evol_parameters}}
\scriptsize
\begin{center}
\begin{tabular}{lcccr}
\hline
\hline
Name&$T_\mathrm{eff}^{\rm LDR}$&$T_\mathrm{eff}^{\rm PHOT}$&$T_\mathrm{eff}^{\rm SPEC}$&[Fe/H]\\
    &     (K)               &         (K)            &       (K)             &       \\
\hline
  HD2114 & 5259 & 5180 & 5288 & $-$0.03 \\ 
  HD2151 & 5973 & 5785 & 5964 & $-$0.03 \\ 
  HD7672 & 5121 & 4957 & 5096 & $-$0.33 \\ 
 HD11977 & 5024 & 4901 & 4975 & $-$0.21 \\ 
 HD12438 & 5236 & 4888 & 4975 & $-$0.61 \\ 
 HD16417 & 5872 & 5729 & 5936 &    0.19 \\ 
 HD18322 & 4639 & 4638 & 4637 & $-$0.07 \\ 
 HD18885 & 4641 & 4673 & 4737 &    0.10 \\ 
 HD18907 & 5143 & 5056 & 5091 & $-$0.61 \\ 
 HD21120 & 5203 & 5026 & 5180 & $-$0.12 \\ 
 HD22663 & 4514 & 4792 & 4624 &    0.11 \\ 
 HD23319 & 4434 & 4526 & 4522 &    0.24 \\ 
 HD23940 & 4962 & 4782 & 4884 & $-$0.35 \\ 
 HD26923 & 5985 & 6126 & 6207 & $-$0.06 \\ 
 HD27256 & 5077 & 5012 & 5196 &    0.07 \\ 
 HD27371 & 4884 & 4914 & 5030 &    0.13 \\ 
 HD27697 & 4901 & 4876 & 4951 &    0.06 \\ 
 HD32887 & 4141 & 4046 & 4131 & $-$0.09 \\ 
 HD34642 & 4775 & 4838 & 4870 & $-$0.04 \\ 
 HD36189 & 5066 & 4875 & 5081 & $-$0.02 \\ 
 HD36848 & 4325 & 4460 & 4460 &    0.21 \\ 
 HD47205 & 4604 & 4777 & 4744 &    0.18 \\ 
 HD47536 & 4513 & 4379 & 4352 & $-$0.68 \\ 
 HD50778 & 4117 & 4081 & 4084 & $-$0.29 \\ 
 HD61935 & 4820 & 4777 & 4879 & $-$0.01 \\ 
 HD62644 & 5590 & 5297 & 5526 &    0.12 \\ 
 HD62902 & 4168 & 4230 & 4311 &    0.33 \\ 
 HD63697 & 4263 & 4346 & 4322 &    0.13 \\ 
 HD65695 & 4483 & 4421 & 4468 & $-$0.14 \\ 
 HD70982 & 5035 & 4957 & 5089 & $-$0.03 \\ 
 HD72650 & 4293 & 4307 & 4310 &    0.06 \\ 
 HD81797 & 4248 & 4085 & 4186 &    0.00 \\ 
 HD83441 & 4575 & 4624 & 4649 &    0.10 \\ 
 HD85035 & 4531 & 4724 & 4680 &    0.12 \\ 
 HD90957 & 4155 & 4121 & 4172 &    0.05 \\ 
 HD92588 & 4972 & 5025 & 5136 &    0.07 \\ 
 HD93257 & 4480 & 4602 & 4607 &    0.13 \\ 
 HD93773 & 5027 & 4912 & 4985 & $-$0.07 \\ 
 HD99167 & 4030 & 3905 & 4010 & $-$0.36 \\ 
HD101321 & 4786 & 4810 & 4803 & $-$0.14 \\ 
HD107446 & 4229 & 4145 & 4148 & $-$0.10 \\ 
HD110014 & 4414 & 4429 & 4445 &    0.19 \\ 
HD111884 & 4306 & 4270 & 4271 & $-$0.06 \\ 
HD113226 & 5027 & 4988 & 5086 &    0.09 \\ 
HD115478 & 4252 & 4293 & 4250 &    0.03 \\ 
HD122430 & 4323 & 4238 & 4300 & $-$0.05 \\ 
HD124882 & 4332 & 4256 & 4293 & $-$0.24 \\ 
HD125560 & 4395 & 4443 & 4472 &    0.16 \\ 
HD131109 & 4154 & 4073 & 4158 & $-$0.07 \\ 
HD136014 & 4999 & 4782 & 4869 & $-$0.46 \\ 
HD148760 & 4564 & 4694 & 4654 &    0.13 \\ 
HD151249 & 4011 & 3885 & 3886 & $-$0.37 \\ 
HD152334 & 4191 & 4162 & 4169 &    0.06 \\ 
HD152980 & 4215 & 4069 & 4176 &    0.01 \\ 
HD159194 & 4383 & 4418 & 4444 &    0.14 \\ 
HD165760 & 4989 & 4929 & 5005 &    0.02 \\ 
HD169370 & 4547 & 4488 & 4460 & $-$0.17 \\ 
HD174295 & 4979 & 4831 & 4893 & $-$0.24 \\ 
HD175751 & 4681 & 4715 & 4710 &    0.01 \\ 
HD177389 & 4991 & 4939 & 5131 &    0.02 \\ 
HD179799 & 4804 & 4879 & 4865 &    0.03 \\ 
HD187195 & 4349 & 4428 & 4444 &    0.13 \\ 
HD189319 & 4091 & 3887 & 3978 & $-$0.29 \\ 
HD190608 & 4647 & 4724 & 4741 &    0.05 \\ 
HD198232 & 4902 & 4824 & 4923 &    0.03 \\ 
HD198431 & 4676 & 4649 & 4641 & $-$0.12 \\ 
HD199665 & 5000 & 4975 & 5089 &    0.05 \\ 
HD217428 & 5214 & 5101 & 5285 &    0.03 \\ 
HD218527 & 5026 & 5066 & 5084 &    0.03 \\ 
HD219615 & 5068 & 4842 & 4885 & $-$0.51 \\ 
HD224533 & 5023 & 4967 & 5062 &    0.00 \\ 
\hline 
\end{tabular}				      
\end{center}
\end{table}
\normalsize

%-----------------------------------------------------------------------------------------------------------------
%------------------------------------------------------------------------------------------------------------------

\subsection{IC 4651}

\subsubsection{Effective temperature}
\label{sec:IC4651_teff}

The star name and effective temperature as derived with the LDR ($T_\mathrm{eff}^{\rm LDR}$) and spectral synthesis 
($T_\mathrm{eff}^{\rm SPEC}$) methods (\citealt{Pasqu04}) are listed in the first three columns of Table 
\ref{tab:IC4651_parameters}. The stars with prefix ``E'' are from \cite{Eggen71}, while the star with prefix ``MEI'' 
is from \cite{Meibom00}.

$T_\mathrm{eff}^{\rm LDR}$ is always lower than $T_\mathrm{eff}^{\rm SPEC}$, by an amount  between 70 and 90~K for 
most of the stars, a difference which is slightly larger than for the field stars. One cause is in the fact that 
IC 4651 is slightly metal rich ([Fe/H]$\sim$0.1, \citealt{Pasqu04}). This implies that we should apply a metallicity 
correction of 19~K to $T_\mathrm{eff}^{\rm LDR}$ to all the stars (cfr. previous Section). With this correction, the 
offset between the two temperatures  becomes of 50--70~K, with the  exception of E95. In the fourth column of Table 
\ref{tab:IC4651_parameters} the values of the effective temperature ($T_\mathrm{eff}^{\rm LDRc}$) derived by the LDR 
method and corrected for metallicity are given.

For the star E95 the difference is much larger, or 320~K. E95 is a subgiant star, as it appears from the CMD in 
Fig.~\ref{fig:cmd_ic4651}. This has two major effects in the present analysis. This is the evolved star for which 
\cite{Pasqu04} found the largest difference between photometric and spectroscopic temperature, and a full convergence 
of the solution could not be obtained. In addition, the LDR value quoted is a weighted average of the values we would 
obtain from the dwarfs calibration ($\sim$5700 K) and the giants calibration ($\sim$5250 K). As previously mentioned, 
we do expect for this star a lower accuracy in its LDR temperature determination. 
 
The presence of only one hot subgiant in the sample does not allow us to derive firm conclusions, but it may indicate 
that before extending the method to stars in this portion of the HR diagram, further work is required. We note, however, 
that the change in temperature of 300 K for this star does not substantially change its derived age and mass, since the 
evolutionary tracks in this portion of the HR diagram are substantially horizontal and the evolutionary phase is extremely 
fast. 

It is more difficult to provide a reliable estimate for the systematic effects which could be present in our results. 
\cite{Carret04}, for instance, have in their sample three stars (namely 27, 76, and 146 of their paper), with magnitudes 
and color indices very close to those of the stars E98 and E60 of our sample. In particular, their star 146 corresponds 
to our E98. Yet, the spectroscopic temperatures they derive (4610, 4620 and 4720, respectively) are 180-290 K cooler than 
what found by \cite{Pasqu04} for E98 and E60, and 130-240 K cooler than derived by us with the LDR method. 

\begin{figure}
\centering
\includegraphics[width=9cm]{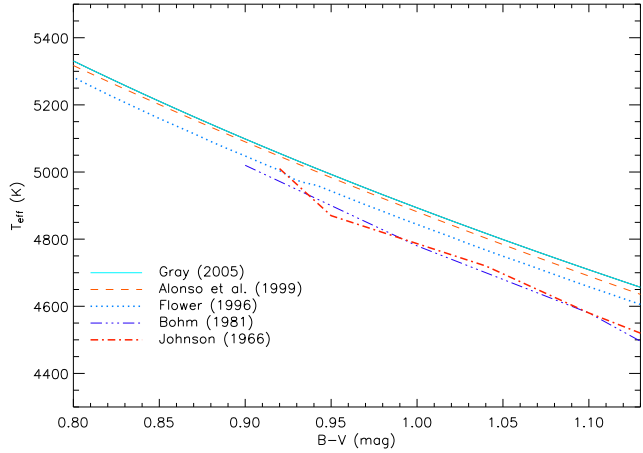}
\caption{Examples of temperature calibrations published by several authors.}
\label{fig:bv_teff_calibr_comparison}
\end{figure}

While a difference on a single star, of say 200 K, can be understood, a systematic effect of up to 200 K with respect 
to the LDR is more difficult to explain. $T_\mathrm{eff}^{\rm LDR}$ for these stars have been obtained on high quality 
spectra and by using a large number of line-depth ratios (namely 15), which makes the measurements extremely robust. As 
far as the calibrations are concerned, our stars occupy the linear part of the LDR$-T_\mathrm{eff}$ relationships, far 
from saturation effects. We therefore consider that the $T_\mathrm{eff}^{\rm LDR}$ measurement errors come mainly from 
the several LDR$-T_\mathrm{eff}^{\rm LDR}$ relationships, which give different temperature values. Any 
systematic shift should therefore be dominated by 
the adoption of different temperature scales. Since our calibrations are ultimately based on the $(B-V)-$temperature 
scale by \cite{Gray05}, we verified the systematic differences in the color range of our interest ($B-V=$0.80--1.13) 
for a number of calibrations present in literature. In Fig.~\ref{fig:bv_teff_calibr_comparison}, we can see that for 
giants of $B-V\sim0.95$ all the scales agree within 100 K, and this agreement increases ($\sim$ 50 K) if we restrict 
ourselves to the most recent ones. We therefore tend to exclude systematic errors in $T_{\rm eff}$ larger than 100 K 
and argue that the temperature determined by \cite{Carret04} as well as the reddening they have adopted are likely too low.

\subsubsection{Surface gravity}
\label{sec:surf_grav_bis}

To derive the photometric surface gravity of the giants studied in IC 4651 using the relation given in 
Section~\ref{sec:surf_grav}, the value of 1.8$M_{\sun}$ suggested by \cite{Pasqu04} as a lower limit 
of the mass of the turn-off stars was assumed, while the photometric temperature $T_\mathrm{eff}^{\rm PHOT}$ was computed 
from the $(b-y)-T_\mathrm{eff}$ calibrations (\citealt{Alonso99}) taking into account the color excess $E(b-y)=0.091$ and 
[Fe/H]=0.10 (\citealt{Pasqu04}). The luminosity $L$ has been derived considering the $V$ absolute magnitude of the 
stars, the bolometric correction ($BC$) tabulated by \cite{Flower96} as a function of $T_\mathrm{eff}$ and the solar 
bolometric magnitude $M_{\rm bol,\sun}=4.75$ (\citealt{Cox00}). 
The absolute magnitude has been computed using the apparent visual distance modulus $V-M_{V}$ determined 
by the best fit of the isochrones at Z=0.024 developed by \cite{Gira00} with the CMD of \cite{Pia98}. For this isochrone 
fitting, the position of the red clump, as well as the main sequence and the turn-off, has been taken into account. This 
is an important point, because theoretical models predict that the absolute luminosity 
of the red clump stars depends fairly weakly on their chemical composition and age. Thus, by the 
fitting of the CMD we find $V-M_{V}=9.83$ for an age of $\log$(Age/yr) = 9.15$\pm$0.05 and $E(B-V)$=0.12. 
This corresponds to an absolute distance modulus of $(V-M_{V})_0=9.46$. The apparent distance modulus we 
find just slightly exceeds the values of 9.7 and 9.8 obtained by \cite{Nissen88} and \cite{Kjel91}, respectively. 

\begin{figure}
\centering
\includegraphics[width=9cm]{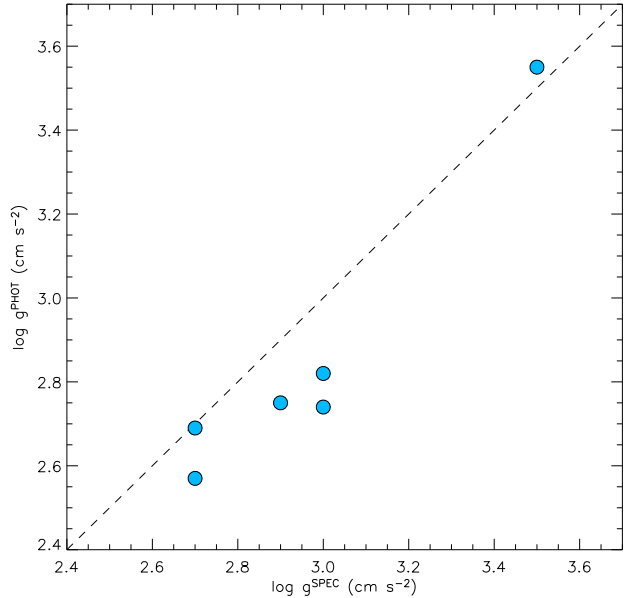}
\caption {Photometric gravities vs. spectroscopic gravities derived by \cite{Pasqu04} for the giant stars studied 
in IC 4651.}
\label{fig:gravity_comparison}
\end{figure}

In Table~\ref{tab:IC4651_parameters} the photometric gravities ($\log g^{\rm PHOT}$) are listed together 
with the spectroscopic ones obtained ($\log g^{\rm SPEC}$) by \cite{Pasqu04}, whose gravities represent a lower limit 
because the \ion{Fe}{i}/\ion{Fe}{ii} ionization equilibrium was not always reached. In 
Fig.~\ref{fig:gravity_comparison} it is shown that the values of $\log g^{\rm SPEC}$ are on average $\approx$ 0.2 dex 
higher than $\log g^{\rm PHOT}$ with differences also of up to 0.3 dex at lower temperatures. \cite{Carret04} justify 
this effect, which they also found for their clump stars in IC 4651, as due to increased \ion{Fe}{ii} line 
blends at lower temperature or to variations of stellar atmospheric structure with temperature. \cite{daSil06} also 
found the same effect for their field stars, where the spectroscopic gravities are systematically overestimated by 0.2 
dex in average, with a dependence on the effective temperature.

%--------------------------------------------------------------------------------------------------------------------
%--------------------------------------------------------------------------------------------------------------------

%--------------------------------------------------------------------------------------------------------------------
%--------------------------------------------------------------------------------------------------------------------

%--------------------------------------------------------------------------------------------------------------------
%--------------------------------------------------------------------------------------------------------------------

\subsubsection{Reddening}

To determine the reddening of IC 4651, we have used the $T_\mathrm{eff}-(B-V)$ calibrations of \cite{Gray05} and 
of \cite{Alonso96, Alonso99}, which includes the metallicity effect ([Fe/H]=0.10$\pm$0.03 for IC~4651, 
\citealt{Pasqu04}). Deriving the intrinsic color index for all the stars from these relationships and comparing them 
with the observed colors, we obtain from the Gray's calibration $E(B-V)=0.111\pm0.018, 0.164\pm0.015$ and from 
the Alonso's calibration $E(B-V)=0.127\pm0.016, 0.167\pm0.017$, in which the two values correspond to the results 
obtained with the two temperatures $T_\mathrm{eff}^{\rm LDR}$ and $T_\mathrm{eff}^{\rm SPEC}$, respectively. 
The color excesses obtained with the two calibrations are very similar and this improves if one considers the effective 
temperatures $T_\mathrm{eff}^{\rm LDRc}$ derived by the LDR method and corrected for metallicity effects. In this 
case the color excesses obtained with $T_\mathrm{eff}^{\rm LDRc}$ is $E(B-V)=0.120\pm0.016$, which is near to the 
value obtained considering the Alonso's calibration. In addition, our color excess values are in good agreement with 
the results of $E(B-V)=0.15$, 0.10, 0.13, 0.10, 0.10 and 0.15 obtained by \cite{Eggen71}, \cite{Nissen88}, 
\cite{Kjel91}, \cite{Meibom02}, \cite{AnthTwa00} and \cite{Pasqu04}, respectively. Since \cite{Meibom02}, 
\cite{AnthTwa00} and \cite{Pasqu04} compute $E(b-y)$, the relation $E(b-y)=0.72E(B-V)$ (\citealt{Cardel89}) has 
been used. These results seem to confirm our analysis of the Section \ref{sec:IC4651_teff} that the reddening value 
derived by \cite{Carret04} of $E(B-V)=0.083\pm0.011$ is likely too low.

\subsubsection{Mass and age}

Since the PDF method allows us to derive complete probability distribution functions separately for 
each analysed stellar parameter, we have been able to derive the mass of the six stars studied in IC~4651 and the 
age of the cluster. 

\begin{figure*}
\includegraphics[width=9.cm]{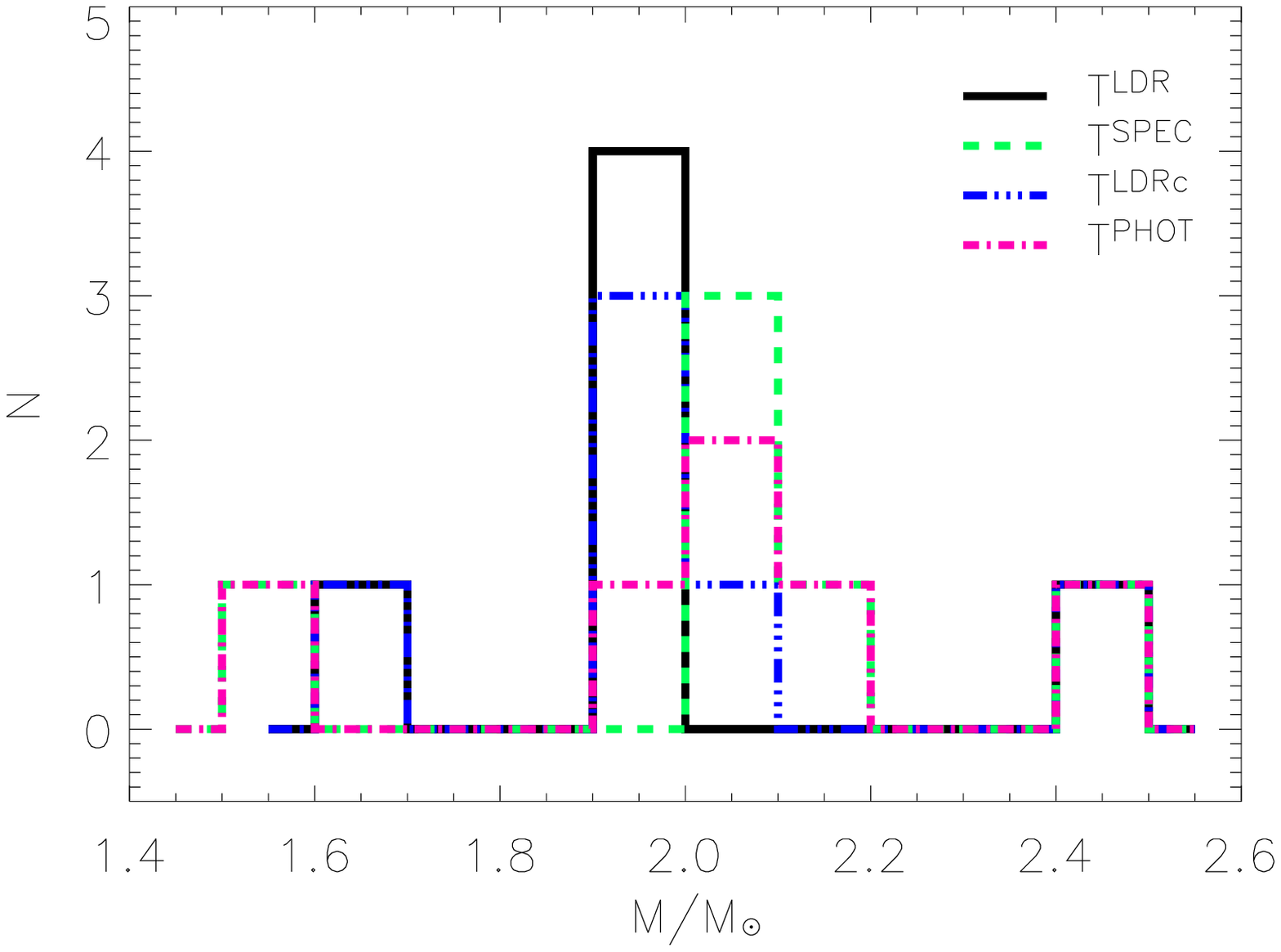}\includegraphics[width=9.cm]{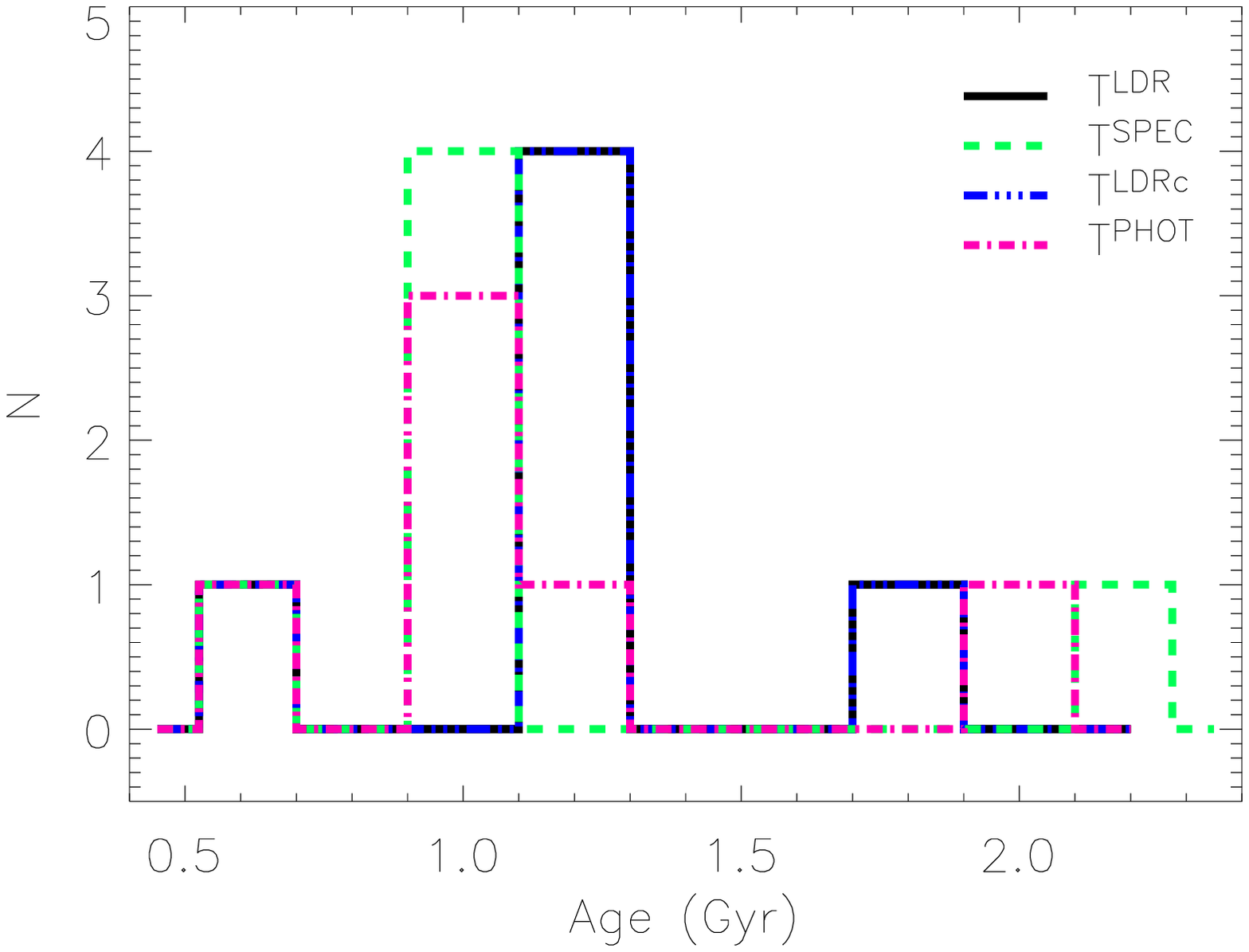}
\caption {Mass and age distributions of the giant stars in IC 4651 for the different values of the effective 
temperature.}
\label{fig:mass_age_distr}
\end{figure*}

The results of the mass and age distributions based on the PDF method are plotted in Fig.~\ref{fig:mass_age_distr} 
and listed in Table~\ref{tab:IC4651_parameters}, both for the temperatures $T_\mathrm{eff}^{\rm LDR}$ and 
$T_\mathrm{eff}^{\rm SPEC}$. The mass and age distributions obtained for $T_\mathrm{eff}^{\rm LDRc}$ and 
$T_\mathrm{eff}^{\rm PHOT}$ are also shown for comparison.

We find that the average mass of the six giant stars studied in IC 4651 is 2.0$\pm$0.2 $M_{\sun}$. \cite{Pasqu04} 
use stellar models of 1.8$M_{\sun}$ to reproduce the lithium abundances as a function of temperature simultaneously 
for ten post-turnoff and giant stars (see their Fig.~5). If we instead consider only the 
six giant stars, the lithium abundance after the dredge-up is $0.2\ltsim\log N(Li)\ltsim2.0$. This 
lithium abundance can be explained by means of a 2.0$M_{\sun}$ rotating model with initial rotational velocity 
$V_{\rm rot}$ of 110 km s$^{-1}$ and with a subsequent modest braking which leads to retain more lithium 
(\citealt{Pala03}). This mass value is consistent with our determination. 

The cluster age found with the PDF method is 1.2$\pm$0.2 Gyr. This agrees with the value of 
$\log$(Age/yr)=9.15$\pm$0.05 ($1.40\pm0.15$ Gyr) that we determined via isochrone fitting (Section~\ref{sec:surf_grav_bis}), 
but is significantly lower than the values of 2.4$\pm0.3$ Gyr and 1.7$\pm$0.15 Gyr found by \cite{Anth88} and 
\cite{Meibom02}, respectively. Discussing the origin of these differences is beyond the scope of this paper. We just 
recall that the main differences between our and their isochrone fitting is that we consider also the position of the 
red clump in the CMD, additionally to the position of the main sequence and turn-off stars. The red clump provides a 
strong constraint to the apparent distance modulus that can be used in the isochrone fitting, thus better constraining 
the cluster age via the turn-off.

Clearly the errors associated to our ages and masses should be intended as internal errors. The fact of using 
different stars average out the errors on the single objects, however, all possible effects due, for instance, 
to a temperature scale shift and to the use of a particular set of evolutionary tracks are not considered. Systematic effects 
will change our results in a different way, depending on the position of the stars in the CMD. For instance, while 
the temperature is not of paramount importance in the Hertzsprung gap (star E95), it is the most critical parameter for 
Red Giant Branch stars. As far as the effective temperature scale is concerned, we tested the PDF method with the three 
stars 27, 76, and 146 studied by \cite{Carret04}, with $T_{\rm eff}=4610, 4620, 4730$ K, $V=10\fm91, 10\fm91, 10\fm94$, 
respectively, and adopting their reddening ($E(B-V)=0.083$) and their apparent distance modulus ($V-M_{V}=$10.15). The 
results are that for the stars 27 and 76 we obtain masses and ages well out of the accepted range (Age$\sim3.4$ Gyr, 
$M\simeq 1.3 M_{\sun}$), while for the star 146 the result (Age$\sim1.7$ Gyr, $M\simeq 1.8 M_{\sun}$) is well compatible 
with our and within the range of the most accepted values avaliable in literature. This shows that while a shift of 
$\sim$100 K doesn't change our analysis in a dramatic way, a difference of 200 K will produce unacceptable results.

%---------------------------------------------------------------------------------------------------------------------------------
%---------------------------------------------------------------------------------------------------------------------------------

\begin{table*}
\caption{Parameters of the giant stars analysed in IC 4651.}
\label{tab:IC4651_parameters}
\begin{center}
\begin{tabular}{lcccccccccc}
\hline
\hline
{Name}&{$T_\mathrm{eff}^{\rm LDR}$}&{$T_\mathrm{eff}^{\rm SPEC}$}&{$T_\mathrm{eff}^{\rm LDRc}$}&{$M_{\rm V}$}&{$\log g^{\rm PHOT}$}&{$\log g^{\rm SPEC}$}&{$M^{\rm LDR}$}&{$M^{\rm SPEC}$}&{Age$^{\rm LDR}$}&{Age$^{\rm SPEC}$}\\
{ }&{(K)}&{(K)}&{(K)}&{(mag)}&{(cm s$^{-1}$)}&{(cm s$^{-1}$)}&{($M_{\sun}$)}&{($M_{\sun}$)}&{(Gyr)}&{(Gyr)}\\
\hline 
E12      & 4926 & 5000 & 4945 & 0.524 & 2.50 & 2.7 & 2.464 & 2.496 & 0.698 & 0.662  \\ 
E8       & 4821 & 4900 & 4840 & 0.867 & 2.63 & 2.7 & 2.018 & 2.156 & 1.197 & 0.999  \\ 
E60      & 4833 & 4900 & 4852 & 1.070 & 2.69 & 2.9 & 2.013 & 2.064 & 1.175 & 1.096  \\ 
E98      & 4806 & 4900 & 4825 & 1.080 & 2.66 & 3.0 & 1.984 & 2.061 & 1.218 & 1.098  \\ 
MEI~11218& 4906 & 5000 & 4925 & 1.259 & 2.75 & 3.0 & 2.027 & 2.069 & 1.115 & 1.044  \\ 
E95      & 5482 & 5800 & 5501 & 2.153 & 3.47 & 3.5 & 1.690 & 1.582 & 1.779 & 2.166  \\ 
\hline 
\end{tabular}				      
\end{center}
\end{table*}

\section{Conclusions}
In this work we have derived accurate atmospheric parameters for field stars and for six giant stars in the open cluster 
IC~4651.

For the former group of stars, with very low extinction and well determined physical parameters 
($T_\mathrm{eff}$, $\log g$, $\xi$), we find a good agreement between the temperatures computed 
by \cite{daSil06} from spectral synthesis and those derived by us with the LDR technique. This gives strong 
support to the effectiveness of the LDR method in deriving effective temperatures.

For the giant stars in the intermediate-age open cluster IC 4651, the LDR method allowed us 
to determine accurate effective temperature values, overcoming the problem of parameters degeneracy 
(in $T_\mathrm{eff}$, $\log g$, $\xi$) encountered in the spectral synthesis method, and to derive the 
reddening of the cluster. We find that our value of $E(B-V)$ is in agreement with previous works 
(e.g., \citealt{Pasqu04}), while it seems it has been underestimated by some other author 
(e.g., \citealt{Carret04}), who likely underestimated the stellar effective temperatures.

Thanks to the PDF method developed by \cite{Jorg05} and slightly modified in 
\cite{daSil06}, we are able to find a mass of 2.0$\pm$0.2$M_{\sun}$ for the giant stars and a cluster age 
of 1.2$\pm$0.2 Gyr, by adopting the effective temperature derived with the LDR technique.

We conclude that our approach is well suitable to derive effective temperatures and reddening of evolved stars and 
clusters with a nearly solar-metallicity. The determination of very precise temperature is of great importance to derive 
stellar mass and age. For this reason, the LDR method could be inverted and used for stellar population studies in alternative 
and/or in addition to those based on photometric data.

\begin{acknowledgements}
The authors are grateful to the anonymous referee for a careful reading of the paper and valuable comments. 
KB has been supported by the Italian {\em Ministero dell'Istruzione, Universit\`a e Ricerca} (MIUR) and by the 
{\em Regione Sicilia} which are gratefully acknowledged.
\end{acknowledgements}

{}

\end{document}